\documentclass[aps,prb,twocolumn,floatfix,superscriptaddress,
showpacs,
amssymb]{revtex4}
\usepackage{graphicx}

\def\ba{\begin{array}}
\def\ea{\end{array}}
\def\be{\begin{equation}\begin{array}{l}}
\def\ee{\end{array}\end{equation}}
\def\bea{\begin{equation}\begin{array}{l}}
\def\eea{\end{array}\end{equation}}
\def\f#1#2{\frac{\displaystyle #1}{\displaystyle #2}}
\def\om{\omega}

\def\de{\delta}
\def\De{\Delta}
\def\va{\varepsilon}

\def\la{\lambda}

\def\bi{\bibitem}
\def\c{\cite}

\def\g{\gamma}

\def\al{\alpha}

\begin{document}
\title{Holstein polarons in a strong electric field:
delocalized and stretched states}

\author{Wei Zhang} \author{Alexander O. Govorov}

\affiliation{Department of Physics and Astronomy, and Nanoscale
and Quantum Phenomena Institute, Ohio University, Athens, Ohio
45701-2979}

\author{Sergio E. Ulloa}
\affiliation{Department of Physics and Astronomy, and Nanoscale
and Quantum Phenomena Institute, Ohio University, Athens, Ohio
45701-2979} \affiliation{Solid State Physics Laboratory, ETH
Z\"{u}rich, 8093 Z\"{u}rich,  Switzerland}

\begin{abstract}

The coherent dynamics of a Holstein polaron in  strong electric
fields is considered under different regimes.  Using analytical
and numerical analysis, we show that even for small hopping
constant and weak electron-phonon interaction, the original
discrete Wannier-Stark (WS) ladder electronic states are each
replaced by a semi-continuous band if a resonance condition is
satisfied between the phonon frequency and the ladder spacing. In
this regime, the original localized WS states can become {\em
delocalized}, yielding both `tunneling' and `stretched' polarons.
The transport properties of such a system would exhibit a
modulation of the phonon replicas in typical tunneling
experiments.  The modulation will reflect the complex spectra with
nearly-fractal structure of the semi-continuous band. In the
off-resonance regime, the WS ladder is strongly deformed, although
the states are still localized to a degree which depends on the
detuning: Both the spacing between the levels in the deformed
ladder and the localization length of the resulting eigenfunctions
can be adjusted by the applied electric field.  We also discuss
the regime beyond small hopping constant and weak coupling, and
find an interesting mapping to that limit via the Lang-Firsov
transformation, which allows one to extend the region of validity
of the analysis.

\end{abstract}

 \maketitle

\section{Introduction}

Quantum electronic transport properties in semiconductor
superlattices in high electric fields have been the subject of
much interest in recent years. Experimental studies have reported
interesting phenomena, such as Bloch oscillations and
Wannier-Stark ({\bf WS}) ladders. \c{exp1}  Moreover, a number of
theoretical predictions have been made, including negative
differential conductivity, \c{con} dynamical localization, \c{dy}
and fractional WS ladders under DC and AC fields, \c{fws} to name
a few.

One important issue for the dynamical behavior of a real system is
the effect of electron-phonon interactions, and extensive research
in this field has been reported.  For example, Ghosh {\em et al}.
and Dekorsy {\em et al}. studied coupled Bloch-phonon
oscillations. \c{blo} The phonon-assisted hopping of an electron
on a WS ladder was studied in [\onlinecite{pho1}]. A brilliant
variational treatment of inelastic quantum transport was given in
[\onlinecite{tru}], and anomalies in transport properties under a
resonance condition were studied in [\onlinecite{res}]. Govorov
{\em et al}. studied the optical absorption associated with the
resonance of a WS ladder and the optical phonon frequency in a
system. \c{gov1,gov2}  A similar and fascinating system of
electron-phonon resonance in magnetic fields has also been studied
extensively. \c{mag}

Much of the work in this area considers incoherent scattering of
electrons by phonons, which is the relevant picture in systems at
high temperature.  In this paper, however, we report a study of
the effects of coherent coupling of an electron to the phonons of
the system, likely to be the relevant description at low
temperatures. We concentrate on how the phonon coupling affects
electron transport in high electric field, in a situation
typically achieved in semiconductor superlattices, for instance,
but also important for electrons in polymer chains, \c{conwell}
and even perhaps reachable in a stack of self-assembled quantum
dots. \c{dots}

Based on a Holstein model, \c{holst} typically used to describe
the small polaron in molecular systems, we analyze the spectrum of
the system and study its transport properties.  We use here a
non-perturbative description, which allows one to elucidate the
effects of resonant and non-resonant phonon fields on the
otherwise localized electrons residing in a WS ladder, for both
weak and strong electron-phonon coupling. We find that for small
hopping and coupling constant, an interesting regime results when
two characteristic energy scales in the system coincide: the one
corresponding to the energy spacing between the WS levels, and the
other corresponding to the frequency of the phonons.  In this {\em
resonant} case, the problem is one of strong mixing between
degenerate states with rather different properties.  The result is
a complex semi-continuous band structure replacing each of the
original WS ladder `rungs', where some of the electron states
become {\em delocalized}, despite the strong electric field
present.  In this case, the phonons interacting with the electron
provide delocalization, unlike the usual situation.  Moreover, and
in contrast to these extended states, `stretched polarons' are
highly degenerate and exhibit strongly localized electronic
components, while the phonon component is in fact extended
throughout the structure. When the system is away from the
resonance condition, a deformed WS ladder will appear, with very
interesting substructure.  The electronic wave functions in this
regime are all localized, but with a localization length that
depends linearly on the degree of detuning.

In all phonon frequency regimes, the rich dynamical behavior of
the system is due to (or reflected in) the near fractal structure
of the spectrum.  This, as we will discuss, has a direct
connection to the Cayley tree structure of the relevant Hilbert
space.  This structure is of course contained in the Hamiltonian
and reflected also in the structure of the eigenvectors.

We also extend our studies beyond the weak coupling and small
hopping constant regime, via a Lang-Firsov transformation.  This
results in an interesting mapping of the problem from the strong
coupling to the weak limit, and allows one to extend the results
for a wider parameter range. A first report of the resonant case
in the weak coupling regime has been published. \cite{EPLus} Here,
we give a detailed theory of such states, including both resonant
and non-resonant cases, and present the strong coupling
interaction regime.

In what remains of the paper, we give in section \ref{description}
a description of the model used and the relevant Hilbert space in
the problem. Section \ref{weak} contains a complete analysis of
the resulting spectrum for small hopping constant and weak phonon
coupling. The extension beyond that regime is given in section
\ref{strong}.  A final discussion is presented in section
\ref{discussion}.

\section{Description of the model} \label{description}

We consider the Holstein model, \c{holst} which describes an
electron in a one-dimensional tight-binding lattice, interacting
locally with dispersionless optical phonons.  Moreover, the system
is subjected to a strong static electric field. The Hamiltonian is
then given by
 \begin{eqnarray}
H_0 &=& \sum_j  \va_j c^+_jc_j+t\sum_j(c_j^+c_{j+1}+c^+_{j+1}c_j)
\nonumber \\
 &+& \om \sum_j a_j^+a_j+\gamma \sum_j c_j^+c_j(a_j^+ +a_j) \, ,
 \end{eqnarray}
where $t$ is the electron hopping constant, $\om$ the phonon
frequency (or energy, with $\hbar=1$), $\g$ the electron-phonon
coupling constant, $\va_j=-edEj\equiv -j \Delta$ the site energy,
$E$ the electric field, and $d$ is the lattice constant.  It is
known that a series of localized WS states will form in a strong
field ($\Delta = edE \gg t$), in the absence of interactions
between electron and phonons.  Within tight-binding theory, the
eigenvalues and eigenfunctions of these WS states are,
respectively $\va_j=-edEj$, and
 \be
 | \phi_j \rangle =\sum_i J_{i-j} (2t/edE) |i \rangle \, ,
 \ee
where $J_{l}$ is the $l$-th order Bessel function.  The WS states
are localized states with characteristic length $2t/eE$.

It is helpful to introduce creation and annihilation operators for
WS states as
 \be
 d_j=\sum_{i=-\infty}^{\infty} J_{i-j}(2t/edE) \, c_i \, .
 \ee
 It is easy to show that this transformation is canonical,
$\{d^+_i,d_j\}=\de_{ij}$, and that the Hamiltonian can be written
in terms of $d_j,d^+_j$ as
 \begin{eqnarray}
 H_0 &=&-\sum_j j\De d_j^+d_j+ \sum_j \om a_j^+a_j \nonumber \\
 ~~ \nonumber \\
 &+& \gamma \sum_{j,i,i'}J_{j-i}J_{j-i'}(a_j+a_j^+)d_i^+d_{i'} \, .
 \end{eqnarray}
In the case of strong electric field (or small hopping constant,
$2t/\De \ll 1$) in which we concentrate in this section, the
Hamiltonian can be simplified as
 \begin{eqnarray}
 H &=& -\sum_j  j\De d_j^+d_j+ \om\sum \limits_j a_j^+a_j
  +\g \sum \limits_j d_j^+ d_j (a_j^+ +a_j) \nonumber \\
  &-& \la \sum_{j}(a_j+a_j^+ -a_{j+1}^+-a_{j+1})(d_j^+d_{j+1}+
   d_{j+1}^+ d_j) \, , \nonumber \\
  \label{H}
 \end{eqnarray}
where the effective coupling constant is now
 \begin{equation}
 \la=\gamma t/\De \, .
 \end{equation}
From (\ref{H}), one can see phonon-assisted hopping between the WS
states quite clearly, so that in fact the phonons introduce
delocalization of the WS electrons. The Hamiltonian (\ref{H}) is
similar to that in [\onlinecite{gov1}], but there are some
differences.  In the current model, when an electron jumps from
one site to the next, it can not only emit (or absorb) a phonon on
(from) its site, but also onto (from) the nearby site. This more
natural description also produces a large difference on the
relevant Hilbert spaces.  The dimension of the relevant Hilbert
space in our case is $2^n -1$, where $n$ is the number of local
sites in the chain, while the size is $n$ for the model in
[\onlinecite{gov1}].  As we will see, the resulting energy
spectrum and other physical quantities show quite a rich behavior.

Let us consider the process of an electron jumping between sites
while creating or annihilating phonons in its neighborhood.  For
example, for a chain of three sites, $n=3$, the hopping of the
electron from the first site will connect the states $|1;000
\rangle$, $|2;010\rangle$, $|2;100\rangle$, $|3;011\rangle$,
$|3;101\rangle$, $|3;020\rangle$, and $|3;110\rangle$.  Here a
vector is expressed as $|j;...m_0,m_1,... \rangle$, where $j$ is
the electron position, and $m_k$ refers to the number of phonons
on site $k$. It is interesting to see that the connectivity of
this portion of Hilbert space has the structure of a Cayley tree,
or a Bethe lattice.\c{car} Under the condition of resonance, i.e.,
$\De= \om$, all the states above for $n=3$ have the same energy.
The off-diagonal matrix elements therefore break the degeneracy
and allow level mixing, and one can expect that some kind of band
may form in the limit of large $n$. As we will see later on, and
in contrast to the resonant case, the spectrum in off-resonance is
still composed of ladders although with a complex structure.
\c{note}

In the basis we list above, the Hamiltonian for $n=3$ takes the
form
\begin{widetext}
 \bea \label{H33}
  H=\left(  \ba{ccccccc}
  -\De & \la & -\la & 0 & 0 & 0 & 0\\
  \la & -2\De+\om & 0  &\la & 0 & -\sqrt{2}\la & 0 \\
  -\la & 0 & -2\De+\om & 0 & \la & 0 & -\la \\
  0 & \la & 0 & -3\De+2\om & 0 & 0 & 0\\
  0 & 0 & \la & 0 & -3\De+2\om & 0 & 0\\
  0 & -\sqrt{2} \la & 0 & 0 & 0 & -3\De+2\om & 0\\
  0 & 0 & -\la & 0 & 0 & 0 & -3\De+2\om
 \ea \right) .
\eea
 \end{widetext}
A direct consequence of the Cayley tree connectivity of the near
degenerate basis is that the Hamiltonian (\ref{H33}) is
constructed by diagonal block matrices for ever larger $n$ (of
sizes 1, 2, 4, ..., 2$^{n-1}$), and mixed by off-diagonal matrix
elements proportional to $\lambda$.  One can see that in the
resonant case, all diagonal elements are given by $-\Delta$,
corresponding to the degenerate manifold. Notice also that the
off-diagonal elements are given by $\pm\lambda$, except for a few
sporadic elements which appear as $\sqrt{2}\la$, associated with
higher number of phonons in a site, such as the state $|3;020
\rangle$ in this case. This nearly self-similar structure appears
for all values of $n$. We expect then that there would be little
change in the physical properties in the large $n$ limit if we
replace $-\sqrt{2}\la $ in those few off-diagonal spots with
$-\la$.  This will be confirmed by our numerical calculation, as
we will see below.  After the substitution, the new {\em
symmetrized} Hamiltonian $H_{sym}$ takes a full self-similar form,
and allows one to exactly solve analytically the eigenvalue
problem, and better understand the physics of the system. Much of
the behavior of $H_{sym}$ remains in the actual system described
by $H$.

\section{Small hopping and weak coupling regime} \label{weak}

Let us consider the eigenvalue problem for the Hamiltonian
$H_{sym}$, defined as that in (\ref{H33}), except that the
elements $-\sqrt{2}\la$ are replaced by $-\la$. By using the block
decomposition formula
 \bea \det \left( \ba{cc}
 A & B\\
 C & D
 \ea \right) = \det(D) \det(A-BD^{-1}C),
 \eea
we find that the eigenvalues of the Hamiltonian $H_{sym}$ are
determined by the equation
 \be \va_0^{N/2} \va_1^{N/4}
 \va_2^{N/8}...\va_{n-1}^1=0 \, ,
 \ee
where $N=2^n$, $\va_0=\va-t_{n}$, $\va$ is the energy eigenvalue,
$t_k$ is defined by
 \be t_k=(n-k)(\om-\De)-\om \, , \ee
and
 \be \va_{k+1}=\va-t_k-\f{2\la^2}{\va_k} \, . \ee
Then, $\va_k$ can be written as a continuous fraction in $k$ steps
 \be
 \va_{k+1}=\va-t_k-\f{2\la^2}{\va-t_{k-1}-\f{2\la^2}{\va -
 t_{k-2}-\f{2\la^2}{...}}} \, .
 \label{contfract}
 \ee

\subsection{Resonant regime ($\De=\om$)}

In this case, $t_k=0$ (after a constant energy shift of $\om$ is
made), we obtain the eigenvalues of $H_{sym}$ as
 \be \va_{k,j}=2\sqrt{2} \la \cos \left(\f{j\pi}{k+1} \right) \, ,
 \label{cos-band}
 \ee
 where $k=1,2,...,n$, and $j=1,...,k$. The asymptotic bandwidth
of the spectrum is therefore given by $4\sqrt{2} \la$.  Notice
that this behavior is similar to that of a system in magnetic
fields. In that case, the electron-phonon interaction breaks the
degeneracy (under the condition of magnetic resonance) and leads
to a ``resonance splitting."  In that case, the splitting is
$\propto \al^{2/3}$ for a three dimensional system, and $\propto
\al^{1/2}$ for a two dimensional system, where $\al$ is the
Frohlich electron-phonon coupling constant. \c{mag}

Let us consider the question of eigenvalue degeneracy by solving
$\va^{N/2}_0 \va^{N/4}_1 \va^{N/8}_2 \va^{N/16}_3=0$, which will
provide us with a good estimate in the large $N$ limit. After some
algebra, one obtains for the eigenvalue equation,
 \begin{eqnarray}
  & & \va^{5N/16} (\va^2-2\la^2)^{N/8} (\va^2-4\la^2)^{N/16}
   \nonumber \\
  & \times & (\va^4 -6 \va^2 \la^2 + 4 \la^4)^{N/16}=0 \, .
  \label{evals}
 \end{eqnarray}
We then have eigenvalues $\va=0$, $\pm\sqrt{2}\la$, $\pm 2 \la$,
and $ \pm \la \sqrt{3 \pm \sqrt{5}}$, with respective degeneracies
of $5N/16$, $N/8$, $N/16$, and $N/16$.  Higher order polynomials
would slightly improve the degeneracy estimate for successively
higher eigenvalues.

Our numerical results for the eigenvalues of $H_{sym}$, Fig.\ 1,
and $H$, Fig.\ 2, exhibit very similar characteristics, as
expected. The different panels in these figures show spectra for
$n=8$, 9, and 11.  One can see that the main structure of the
spectra changes little with increasing $n$, although finer
structure appears for higher $n$, filling to some extent the gaps
of the previous structure. Our analytical results match exactly
those of Fig.\ 1.  The self-similar structure of the spectrum is
the manifestation of such symmetry in $H_{sym}$.  One can see that
the original highly degenerate manifold is broadened into a
semi-continuous band by the off-diagonal mixing elements of
$H_{sym}$. Notice however that large residual degeneracies remain
at the center of the band, and other symmetrical values, as
described above by Eq.\ (\ref{evals}).

The structure of the full Hamiltonian $H$ is shown in Fig.\ 2,
which still maintains a nearly self-similar structure, despite the
sporadic $- \la \sqrt{2}$ `asymmetrical' terms.  Notice however
that the degeneracy at $\va =0$ and other `plateau values' is not
exact here, but only seen as a slight break (or slope) of each
plateau.

\begin{figure}
\includegraphics[width=3.4in]{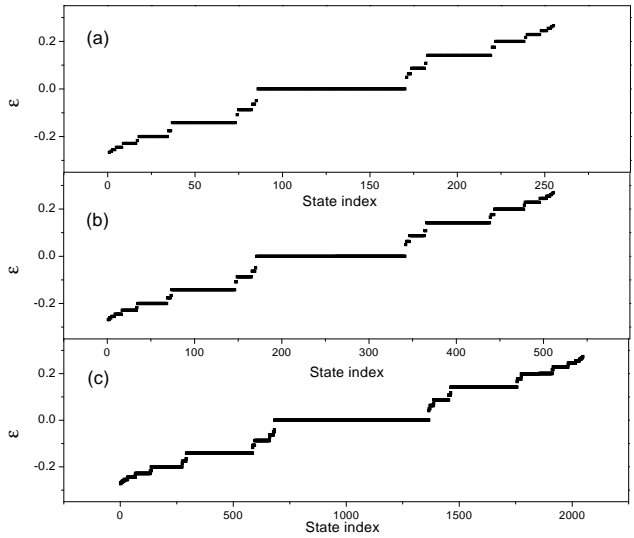}
\caption{ Energy spectra for fully self-similar $H_{sym}$ for
systems with different lattice size, (a) $n=8$, (b) $n=9$, and (c)
$n=11$. Notice the clear fractal structure of spectra, as
anticipated from the structure of $H_{sym}$.  Increasing $n$
produces finer scale structure in the gaps. Here $\De=\om=1$, and
$\la=0.1$, so that the asymptotic bandwidth is $4\sqrt{2} \la =
0.5657$.}
 \label{fig1}
\end{figure}

\begin{figure}
\includegraphics[width=3.4in]{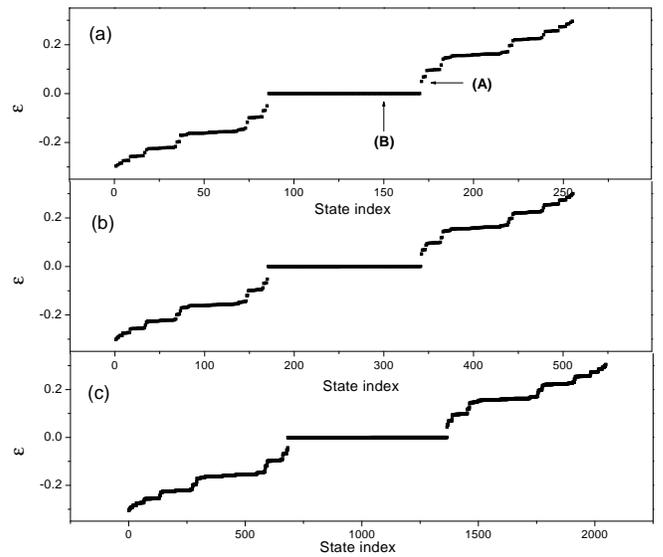}
\caption{Energy spectra for $H$ of systems with different lattice
size, (a) $n=8$, (b) $n=9$, and (c) $n=11$. Notice the approximate
fractal structure of the spectra.  Increasing $n$ gives similar
structure.  Here $\De=\om=1$, $\la=0.1$, and the bandwidth
slightly increases from one panel to the next. A and B labels in
panel (a) indicate states described in Fig.\ \ref{fig4} and
\ref{fig5}.}
 \label{fig2}
\end{figure}

\begin{figure}
\includegraphics[width=3.9in]{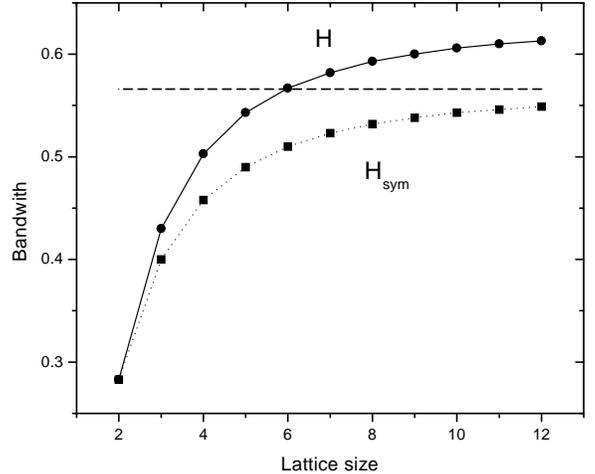}
\caption{Bandwidth results for $\la =0.1$ and different values of
$n$.  The asymptotic bandwidth for $H_{sym}$ is
$4\sqrt{2}\la=0.5657$, shown as dashed line.  We see that
bandwidth for $H$ is $W \agt 6 \la$.}
 \label{bandwidths}
\end{figure}

Figure \ref{bandwidths} compares the results for the bandwidth of
$H$ and $H_{sym}$ from both analytical and numerical results.
Notice that $4\sqrt{2}\la$ is the asymptotic (large $n$)
analytical result from Eq.\ (\ref{cos-band}), for $H_{sym}$.  We
see in Fig.\ \ref{bandwidths} that the numerical results for the
bandwidth of $H_{sym}$ converge quickly to the analytical
prediction.  Similarly, the numerically obtained bandwidth of $H$
is only slightly larger than the bandwidth of $H_{sym}$, and
clearly has also a finite asymptotic value. This is an interesting
feature of $H$, that although one has large Hilbert space
dimension $N-1=2^n-1$ (= 2047 for n=11, for example), one obtains
a finite bandwidth. This is of course also associated with the
fact that there are large degeneracies in the energy spectra,
which increase with $N$ (see Eq.\ \ref{evals}).

\begin{figure}
\includegraphics[width=3.4in]{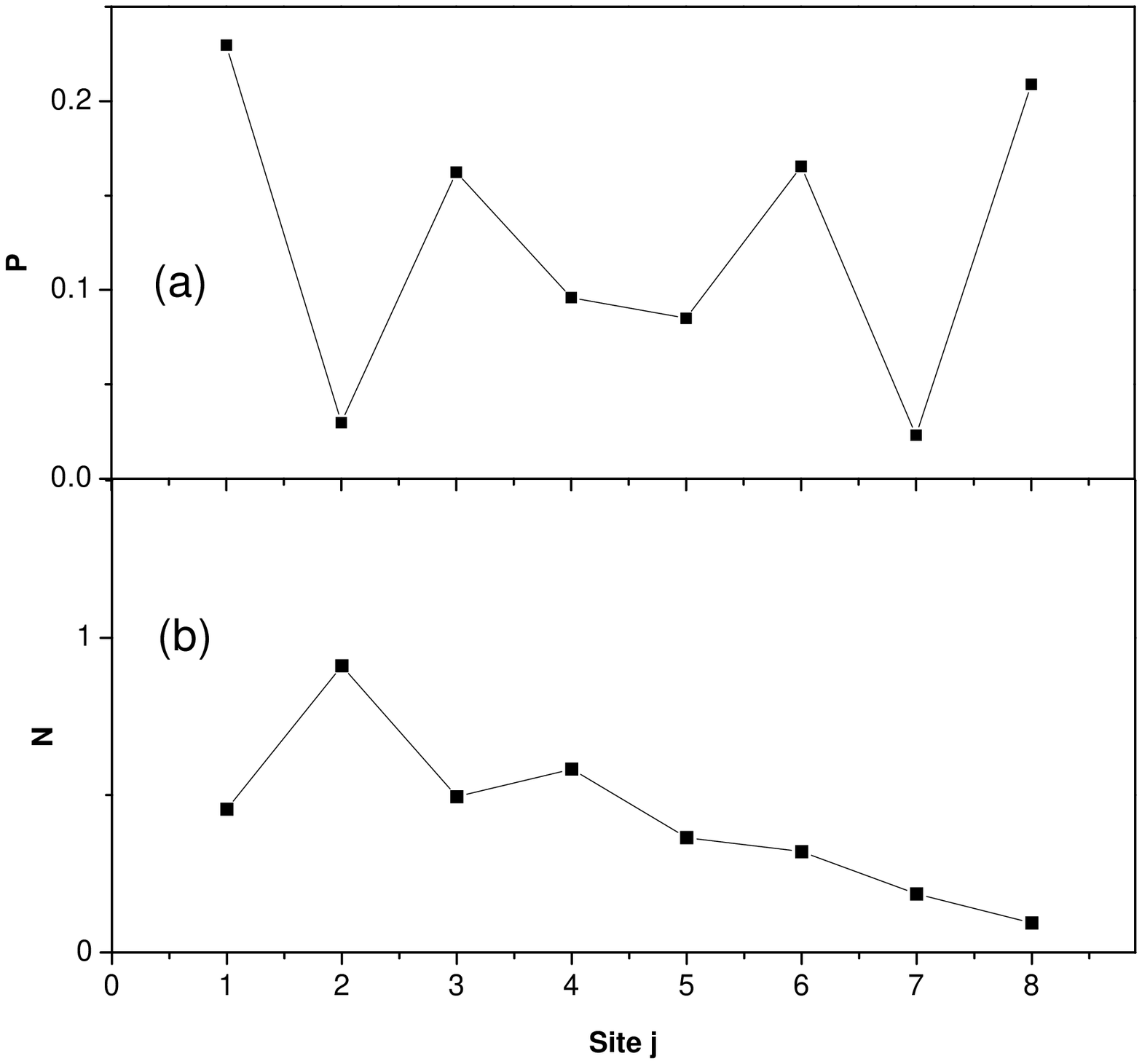}
\includegraphics[width=1.3in,angle=0]{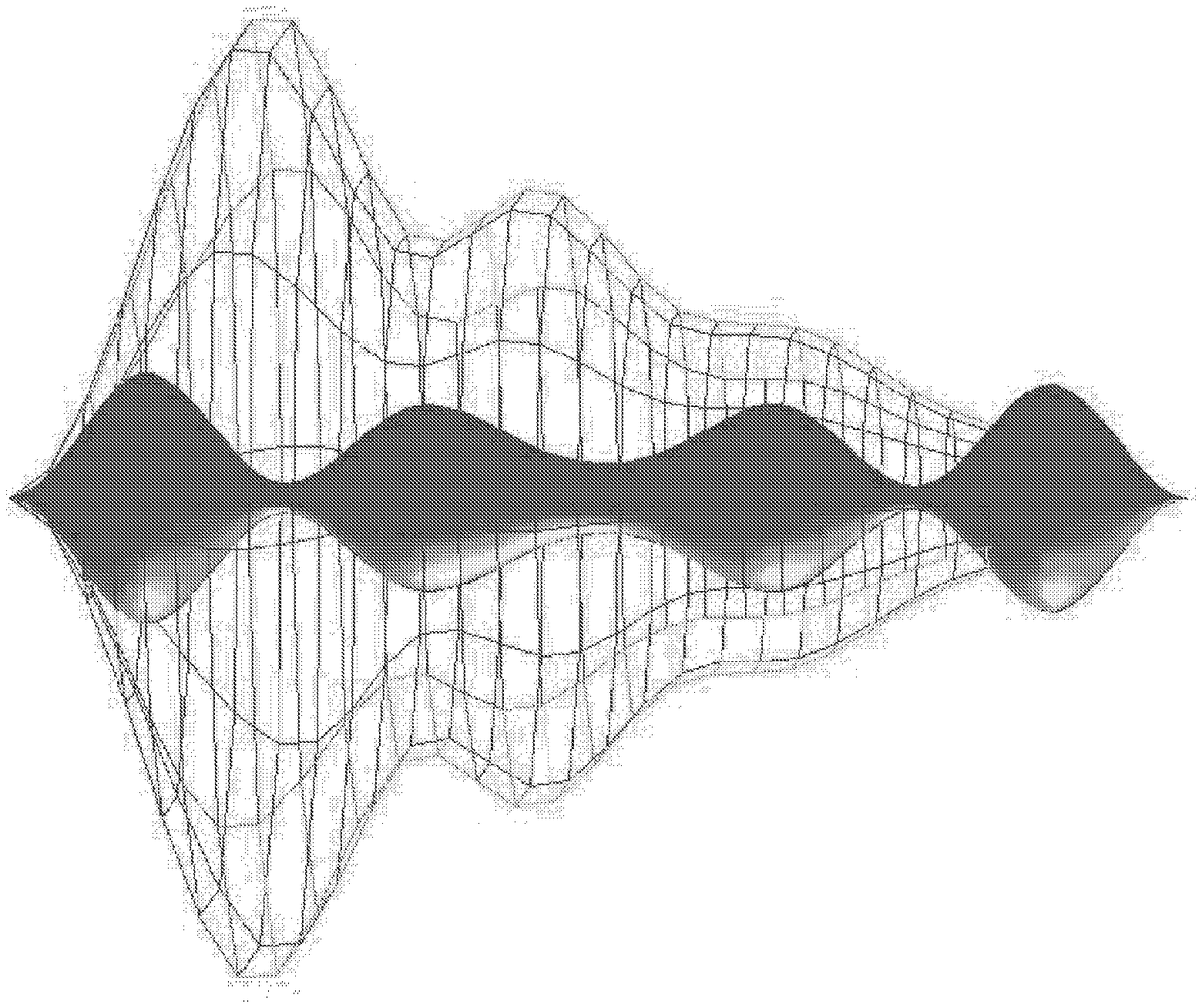}
\caption{(a) Electron probability $P(j)$ for each site $j$, and
(b) spatial distribution of phonons $N(j)$ for state labelled A in
Fig.\ 2a.  Large electron amplitude and phonon number throughout
describe an extended, delocalized polaron, jointly illustrated in
the composite figure at bottom.}
 \label{fig4}
\end{figure}

\begin{figure}
\includegraphics[width=3.4in]{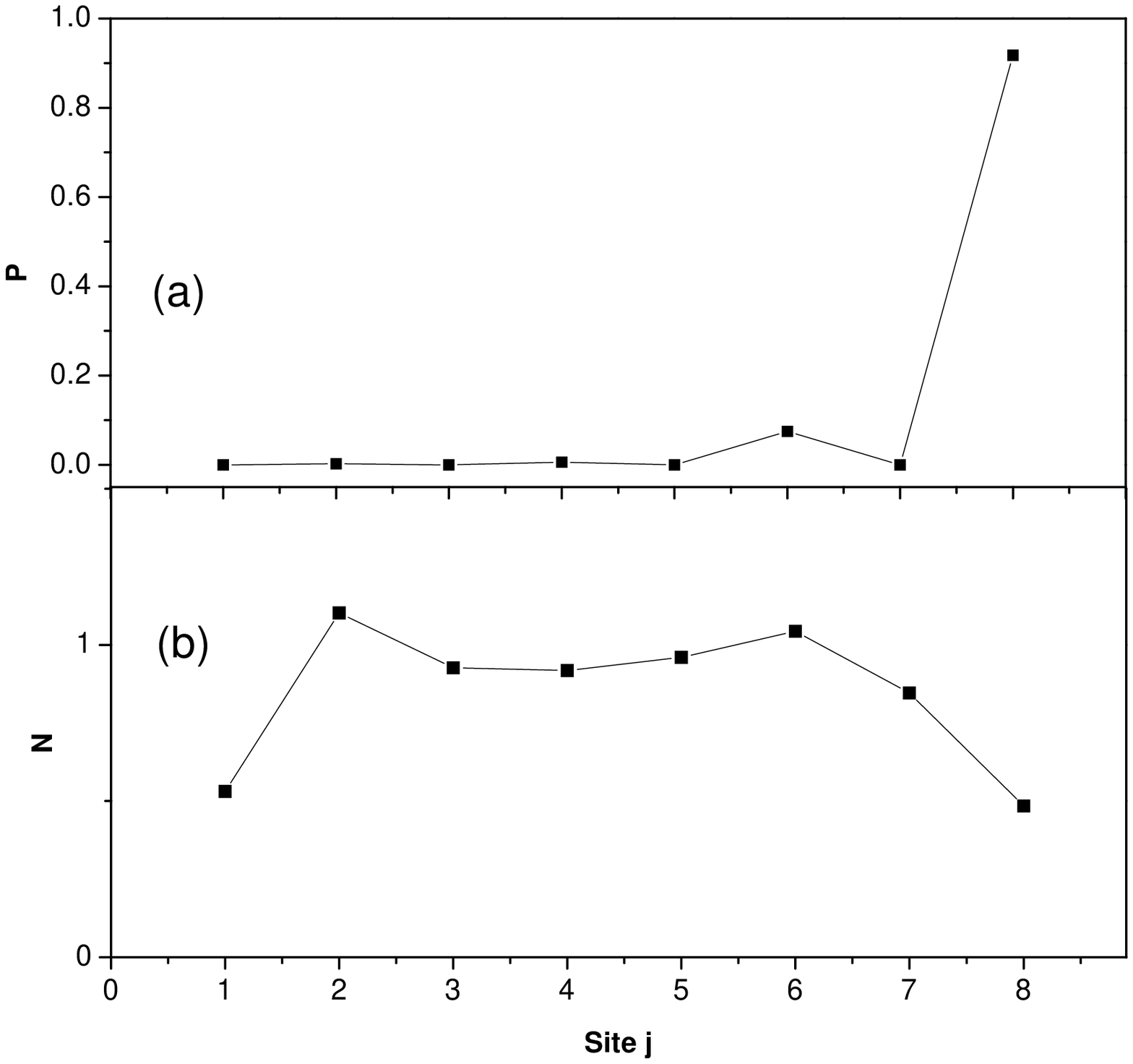}
\includegraphics[width=1.3in,angle=0]{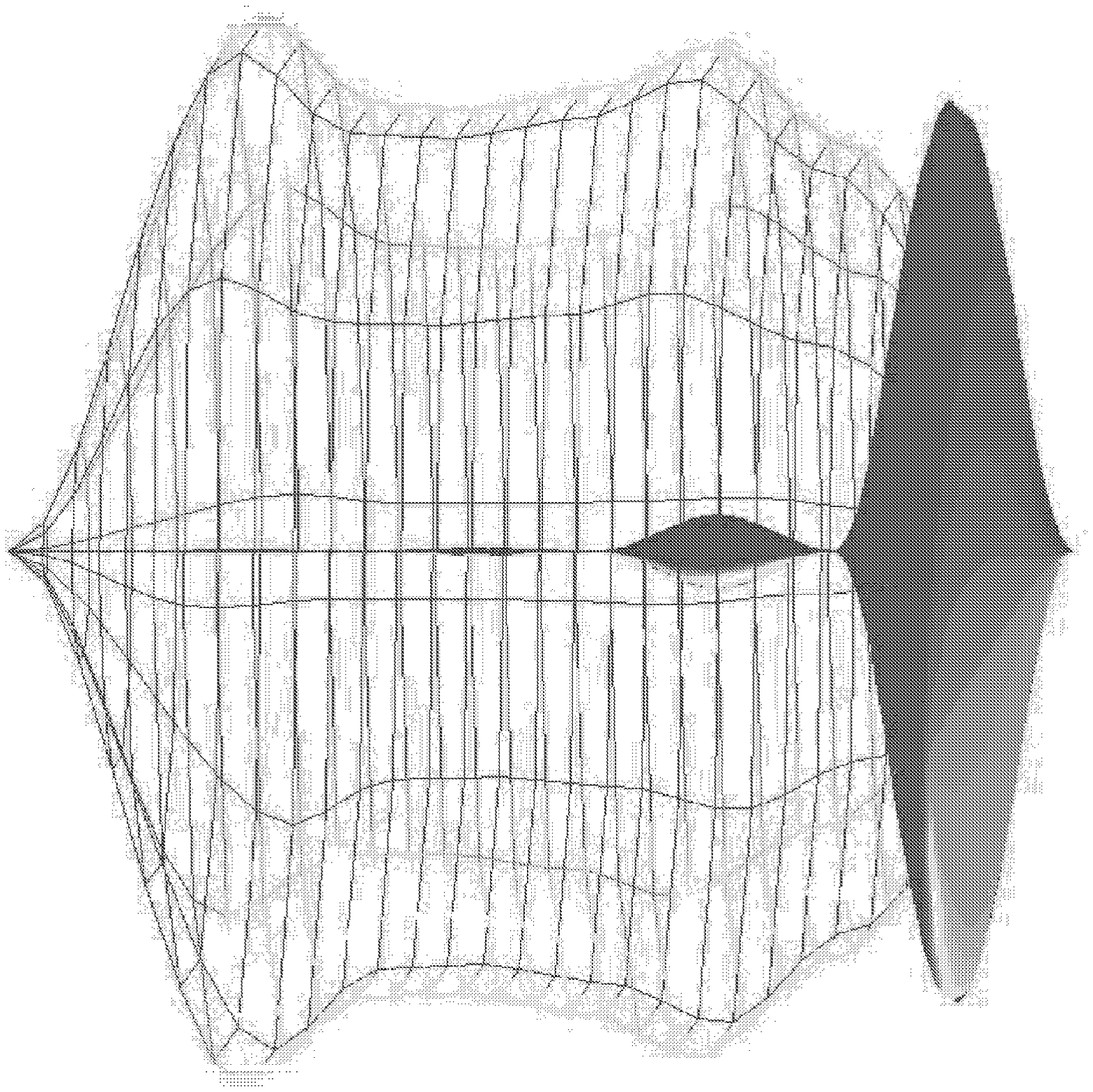}
\caption{(a) Electron probability $P(j)$ for each site $j$, and
(b) spatial distribution of phonons $N(j)$ for state labelled B in
Fig.\ 2a.  Electron is localized at right end of the chain, while
phonons lie throughout, yielding a stretched polaron, illustrated
at bottom figure. }
 \label{fig5}
\end{figure}

To get better understanding of the physics, we show characteristic
electron probabilities, as well as phonon spatial distributions,
for different eigenstates in Fig.\ \ref{fig4} and \ref{fig5}.  The
electronic probability function is given by
 \be
 P_\nu (j) = \langle \Psi _\nu | d^+_j d_j | \Psi _\nu \rangle =
 \sum _{\{m\}} |C^\nu _{j,\{m\}} |^2 \, ,
 \ee
 where the coefficients in the last expressions are obtained from
the diagonalization of $H$, so that the eigenvectors are given by
 \be
 | \Psi _\nu \rangle = \sum _{j,\{m\}} C^\nu _{j,\{m\}} \,
 |j;\{m\} \rangle \, ,
 \ee
 for each $\nu$-eigenstate.  Similarly, the corresponding phonon
spatial distribution is given by
 \be
 N_\nu (j) = \langle \Psi _\nu | a^+_j a_j | \Psi _\nu \rangle =
 \sum _{l,\{m\}} |C^\nu _{l,\{m\}} |^2 m_j \, ,
 \ee
 for each eigenstate.  These two spatial distribution functions
give us an idea of how the two different components of each state
are related to one another.  These functions are a projection of
the rather subtle coherent interactions (or mixtures) between the
electron and phonon subsystems.  One can see in Fig.\ 4a that for
the non-degenerate state A in Fig.\ 2a, the electron is extended
throughout the $n=8$ lattice.  At the same time, the phonon
amplitude is also extended along the lattice, and one can picture
the phonon `cloud' as `surrounding' the electron all along in
Fig.\ 4b, effectively describing a `tunneling polaron'. In
contrast, the degenerate state B at $\va =0$ in Fig.\ 2a, has its
electron component highly localized at the right end of the
lattice, Fig.\ 5a, while the phonon cloud is away, Fig.\ 5b, and
nearly `detached' from the electron.  One can describe this as a
`stretched polaron' (a precursor of the polaron dissociation
predicted at high fields in polymer systems \c{conwell}).

Although by definition the electronic distribution function
$P_\nu$ is normalized to unity, the phonon function $N_\nu$ is not
necessarily so, since the electron can create and absorb phonons
as it propagates up and down the lattice.  This difference in
phonon content for each state can be seen by comparing Fig.\ 4b,
and 5b, as the {\em total} phonon number is clearly larger in the
latter (stretched polaron) case.  To study this feature throughout
the spectrum, we show in Fig.\ 6 the total average phonon number
of all states in the chain $n=8$. One can see clearly that states
at the center of the band ($\va \approx 0$) have many more phonons
($\approx 7=n-1$) than the rest.  An estimate for the average
number of phonons for different states can be obtained by the
following argument.  The average phonon number is given by
$\langle N_\nu \rangle=\sum_i P_\nu (i) m_i$. For an extended
state, $P_\nu \simeq 1/n$, for a chain of $n$ sites. A state with
an electron at site $i$ can be obtained by the electron hopping
$i-1$ steps from the first site, while emitting $m_i \approx i-1$
phonons in the process.  For extended states, this yields $\langle
N \rangle \simeq \sum_i (i-1)/n=(n-1)/2$, which matches well the
numerical results in Fig.\ 6, where the most extended states have
$\langle N \rangle \simeq 3.5$ (such as the state labelled A). In
contrast, an electron localized at site $n$, as B in Fig.\ 6, has
emitted $n-1$ phonons in the process, just as one finds in the
exact calculation.

\begin{figure}
\includegraphics[width=3.4in]{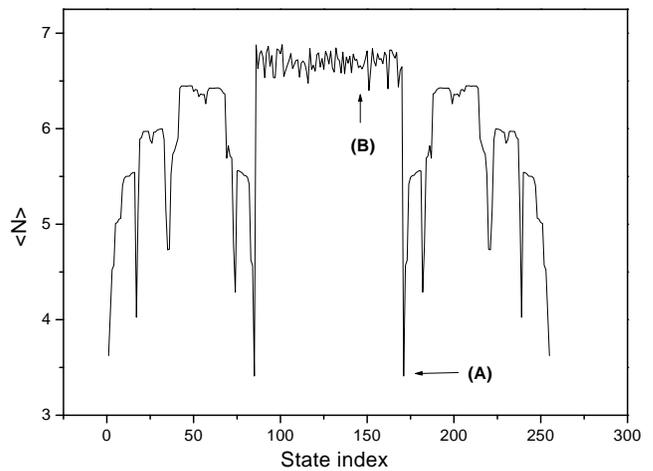}
\caption{Total average phonon number for each eigenstate in Fig.\
2a. Higher value plateaus are associated with the more localized
electronic amplitudes, such as state B.  Lowest phonon counts
correspond to extended electronic states, such as A.}
 \label{fig6}
\end{figure}

\subsubsection{Transport through structure}

We should emphasize that the properties illustrated in Fig.\ 4 and
5 are quite generic: states in one of the highly (nearly)
degenerate levels show different degrees of electron localization
and an abundance of phonons, in a stretched polaron configuration.
In contrast, states with non-degenerate companions are delocalized
polarons throughout the chain with low phonon content.  Even
though is the presence of phonons which delocalizes the electron
(in the sense of the original WS ladder), we obtain that the
localized electrons are accompanied by many phonons, reminiscent
of the self-localized polarons when the coupling is strong.

The rather complicated behavior of the states would of course be
reflected in various properties of the system.  Let us focus here
on what one could measure if electrons where injected from one end
of the chain and were collected out of the other end.  This is
motivated by the ability to carry out just such experiments under
strong electric fields in semiconductor superlattices, as well as
in other systems.  The transport properties provide information
about the density of states in the structure and the spatial
charge distribution of the different states (for optical response,
see [\onlinecite{EPLus}]). The contribution of various eigenstates
to the charge transport can be described by the quantity
 \be
 D_\nu =P_{\nu}(1) P_{\nu}(n) \, ,
 \ee
since the tunneling amplitude is proportional to the density of
states in the leads and the wavefunction amplitudes at both ends
of the structure (the site $j=1$ and $j=n$). \cite{tul} In fact,
the tunneling probability can be calculated from the $S$-matrix
formalism as $|T|^2$, where $T=\langle \va_f,R|S|\va_i,L \rangle$.
Here $R$ and $L$ refer to the right and left leads. In the wide
band limit,
 \begin{eqnarray}
 T & \propto & \Gamma \int dt_1 dt_2e^{i(\va_f t_2-\va_i t_1)}
 \langle n|G^{R}(t_2-t_1)|1\rangle \nonumber \\
 &=& \sum_\nu \Gamma \langle n|\nu\rangle \langle
 \nu|1\rangle \de(\va_f-\va_i) \, ,
 \end{eqnarray}
where $\Gamma$ describes the electron interaction with contacts,
and $G^{R}$ is the retarded Green function connecting both ends of
the structure ($j=1$ and $j=n$).  We see then that the quantity
$D_\nu = |\langle n| \nu \rangle \langle \nu |1 \rangle |^2
\propto |T|^2$ is directly involved in the tunneling probability.
Figure 7a shows the density of states and 7b shows the quantity
$D_\nu$ for the system with $n=8$.  We can see that {\em all} the
states at the center of the band contribute {\em zero} to the
transport amplitude through the chain, and this behavior is
exhibited by all the high peaks in the DOS.  This is clearly
consistent with the spatially localized charge nature of the
highly degenerate states. On the other hand, $D$ shows large
values for states `in the gaps', confirming in fact that the
non-degenerate states in the spectrum have an extended nature. The
variations shown in $D$ would then be reflected in strong
amplitude modulations within each of the phonon replicas in
tunneling experiments, \cite{exp2} whenever the resonance regime
is reached.

\begin{figure}
\includegraphics[width=3.4in]{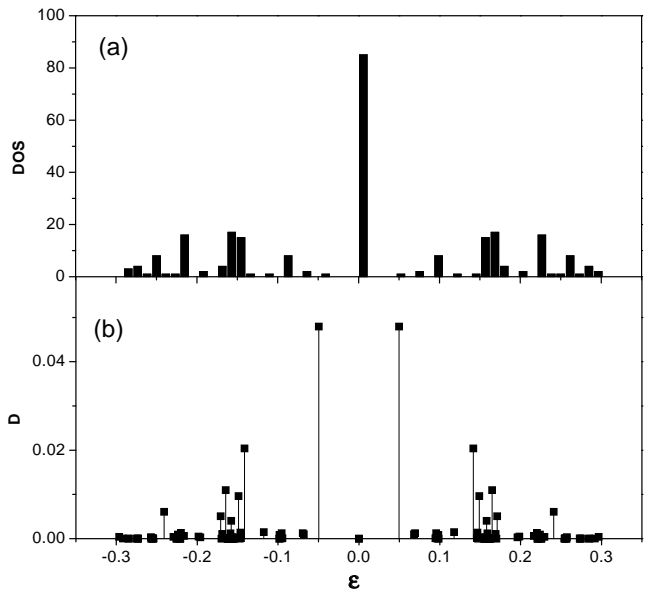}
\caption{ (a) Density of states, and (b) tunneling probability
function $D_\nu$ for $n=8$ system. }
 \label{fig7}
\end{figure}

\subsection{ Non-resonant regime ($\Delta \neq \omega$)}

By studying the equivalent continuous fraction (\ref{contfract})
for the non-resonant regime, we find that the spectrum for this
case is quite different from that under resonance.  For the case
of small coupling constant, i.e., $\la /\om \ll 1$, and $\la / \De%
\ll 1$, the unperturbed spectrum is a series of ladder levels
$\va_j=-(\om-\De)j$ ($j=1,...,n$) instead of the degenerate
manifold in the degenerate case. Notice that the spacing between
the ladders is given by the detuning, $\De' \equiv |\om-\De|$,
instead of the original WS ladder energy.  The correction
introduced by the interaction breaks some of the detuned ladder
degeneracies and gives some substructure to the ladder.  Solving
the first three factors, $\va_0^{N/2}\va_1^{N/4}\va_2^{N/8}=0$, to
second order, $\la^2$, we find that the spectrum is a series of
states with energies $n\De'$, $n\De' + 2\la^2 /\De'$, $(n-1)\De'$,
$(n-1)\De' - 2\la^2 /\De' $, etc., with respective degeneracies
$N/4$, $N/4$, $N/8$, $N/8$, etc.  Notice that here the
electron-phonon produces level splittings $\sim \la^2$, unlike the
resonant case where the splitting (bandwidth) is linear in $\la$.
The problem is also solved numerically and we show the resulting
spectra in Fig.\ 8, for both $H$ (Fig.\ 8a), and $H_{sym}$ (Fig.\
8b).  Our analytical results match the numerical results for
$H_{sym}$ quite well.  We notice again that there are only small
differences between the spectra of $H$ and $H_{sym}$.

\begin{figure}
\includegraphics[width=4.6in]{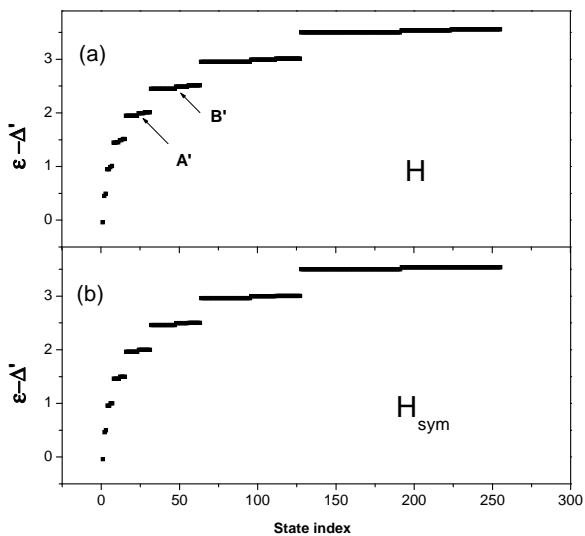}
 \vspace{-0.5in}
\caption{Energy spectra for system with lattice size $n=8$. (a)
For full Hamiltonian $H$, and (b) for $H_{sym}$. Here
$2\De=\om=1$, and $\la=0.1$. A' and B' states are shown in Fig.\
9.}
 \label{fig8}
\end{figure}

Figure 9a, shows the electron probability for different states in
the spectra, labelled A' and B' in Fig.\ 8a.  Figure 9b shows the
distribution of phonons for the same states. It is apparent that
the electron wave function is localized, and separated from its
phonon cloud.  The two states only differ, basically, on which
site to be localized about (as one would expect from their
position in different `rungs' in the ladder in Fig.\ 8a).  In
fact, the different states in the $\De'$ ladder are associated
with different lattice sites (the highest rung states at $\va
\simeq n\De'$ being localized at $j \simeq n$). We also show in
Fig.\ 10 the average phonon number per state for the entire
spectrum. The structure of this figure is quite different from
that in Fig.\ 6, as they reflect dramatically different dynamical
behavior.  A phonon counting argument as the electron hops and
emits phonons can explain the average number of phonons in the
different states in Fig.\ 10, although clearly here it produces a
nearly negligible mixing of states ($\sim \la^2$).

\begin{figure}
\includegraphics[width=3.8in]{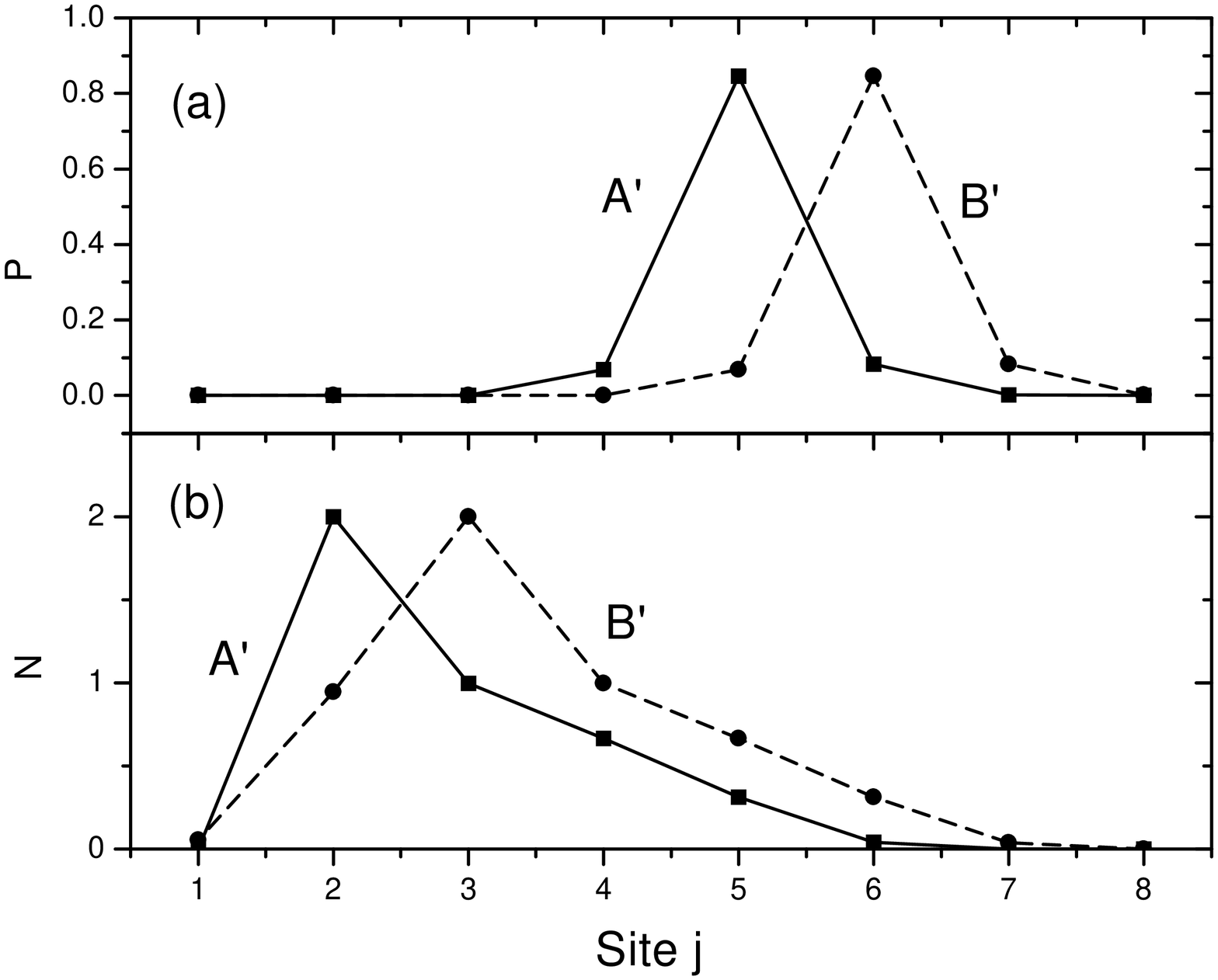}
\caption{(a) Electron probability $P(j)$ for each site $j$ for
states A' and B' indicated in Fig.\ \ref{fig8}a. (b) Corresponding
phonon distributions $N(j)$. }
 \label{fig9ab}
\end{figure}

\begin{figure}
\includegraphics[width=4.2in]{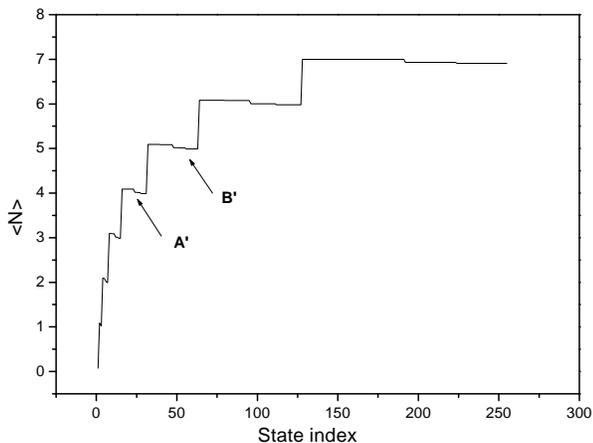}

 \vspace{-1in}
\caption{Total phonon number for each eigenstate in the
non-resonant regime, as in Fig.\ \ref{fig8}.}
 \label{fig10}
\end{figure}

One can further explore the extent of the localization nature of
the eigenstates. In Fig.\ 11, we show the electronic probabilities
of state (A') for various $\De'$ (or electric field) values. We
might expect that the localization length would be $2 \la / \De
'$, like the localization length for the original WS state is $2t/
\De $.  This simple expectation is confirmed by numerical
calculations, as seen in Fig.\ 12, where the solid line shows the
$2\la/\De '$ dependence, and the dots indicate the localization
length for states of various $ \De '$ [the localization length $L$
is extracted from a fit to an exponential amplitude drop about its
central site, $e^{-|x|/L}$].  One would in fact expect that the
manifestation of localization in real space (and energy) would be
susceptible to be measured via Bloch oscillations, just as in a
typical WS ladder. In this case, however, the frequency of Bloch
oscillations is $\De'$, and not $\De$.  Thus the electron-phonon
interaction changes the frequency of Bloch oscillations, and this
effect may be possibly observed in experiment.

\begin{figure}
\includegraphics[width=3.3in]{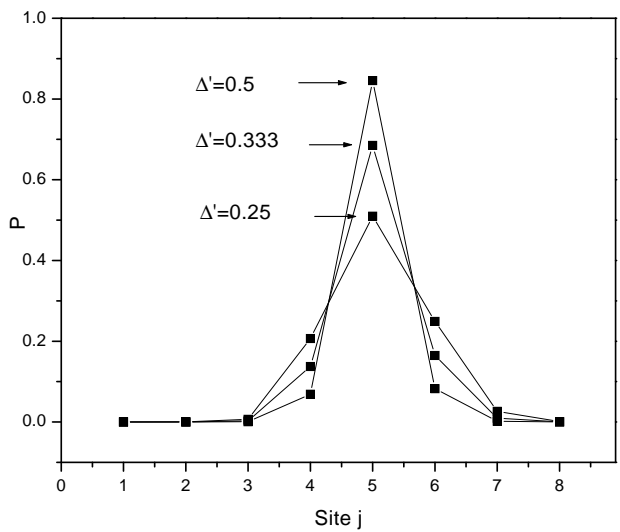}
\caption{Electron probability $P(j)$ for state A' in Fig.\
\ref{fig8}a for various detuning values $\De'$. }
 \label{fig11}
\end{figure}

\begin{figure}
\includegraphics[width=3in]{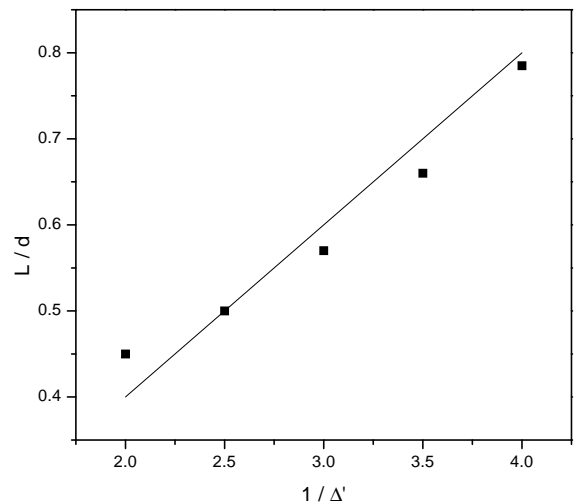}
\caption{ Relation between localization length $L$ and $1/\De'$.
Solid line shows $L/d=2\la/\De'$ dependence, while the dots show
value of $L$ obtained from fit of functions in Fig.\ \ref{fig11}
to the form exp$(-|x|/L)$. }
 \label{fig12}
\end{figure}

From the discussion, it is clear that the eigenstates are
localized in the non-resonant regime.  As such, they contribute
little to transport across the chain.  One can easily calculate
the quantity $D_\nu$ as before.  As expected, the value of $D_\nu$
is very small ($< 10^{-12}$) for every state, and very different
from the resonant case.  This behavior (not shown), is completely
consistent with the localized nature of the eigenstates.  Thus, we
see that in this case the dominant conduction behavior, if any,
would be thermally activated hopping conduction, instead of band
conduction.

\section{Beyond small hopping and weak coupling regime}
\label{strong}

All the discussions so far have been focused on the Hamiltonian
$H$ (or $H_{sym}$), valid only under the condition of small
hopping constant $t/\De \ll 1$.  In this section we extend our
discussion beyond this regime, which in principle may be relevant,
depending on different system parameters.

We use the Lang-Firsov canonical transformation $H^*=e^{S} H_0
e^{-S}$, where $S=-g\sum_j c_j^+c_j(a_j-a_{j+1})$, and $g=\g/\om$.
The transformed Hamiltonian takes the form
 \begin{eqnarray}
 H^* &=& \sum_j \va_j c_j^+c_j+\om \sum_ja_j^+a_j-g^2\om c_j^+c_j
 \nonumber \\
 &+& te^{-g^2}\sum_j [ c_{j+1}^+ c_j e^{g(a^+_{j+1}- a_j^+)}
 e^{-g(a_{j+1}-a_j)} +h.c.] \, .  \nonumber \\
 \end{eqnarray}
One can see that this Hamiltonian is suitable for studying strong
coupling dynamics. As in the case of $H$, we use the operators for
WS ladder states, $d^+_j$. Then the Hamiltonian can be rewritten
in the form
 \begin{widetext}
 \be
 \tilde{H} = -\sum_j (g^2+j)\De d_j^+d_j+\om \sum_j a_j^+ a_j
 + gt'J^2_o(2t'/\De) \sum_j[(a_{j+1}-a_{j+1}^+-a_j+a_j^+)
 (d_j^+d_{j+1}-d_{j+1}^+d_j)] \, ,
 \ee
 \end{widetext}
where $t'=te^{-g^2}$.  We see then that the effective hopping
constant $t'$ becomes smaller than $t$. Thus the problem can be
mapped to that in the previous section, with effective hopping
constant $t'$, and effective coupling constant
$\la'$,
 \be
 t'=te^{-g^2} \, , \quad{} \quad{} \la'=gt'J_0^2(2t'/\De) \, .
 \ee
 It is clear that $\la' \rightarrow 0$, when $g \rightarrow 0$, as
expected, and the spectrum is the discrete WS ladder in the
absence of electron-phonon interaction.  Moreover, one can see
that for weak coupling, the bandwidth $W \simeq 6 \la'$ is
proportional to $g$, just as in the situation discussed in III.A
(see Fig.\ 3).  Beyond the weak coupling and small hopping regime,
however, $W$ and $\la'$ have a nonlinear dependence on $g$. Figure
13 shows the coupling constant dependence of the bandwidth $W$ for
the resonant and stronger coupling limit, $\De = \om = t =1$.
Notice that the bandwidth $W \simeq 1$ for $\gamma \agt 0.6$. At
this coupling, the bands in neighboring WS rungs will begin to
overlap (and interband terms would need to be considered). It is
clear that these corrections to the dependence of $W$ appear also
as function of $t$, $\Delta$, and/or $\om$.

Similar discussions are also relevant in the off-resonant case.
For example, since the localization length is proportional to
$2\la' /\De'$ (section III.B), there is also a correction to the
simple relation $L \propto 2\la / \De'$, when $t$ and/or $g$ are
not small.  For example, with increasing coupling constant, the
correction to the ladder $\sim 2{\la'}^2/\De'$ may become
comparable with the spacing between the detuned ladder levels
$\De'$.  In that case, the ladder structure for the off-resonance
case will also disappear.  In this situation, the electron can
also become delocalized, even if in the non-resonant regime, but
due to the strong coupling $\gamma$.

It is interesting to compare our results to those of Bonca and
Trugman,\c{tru} as one goes from the weak to the strong tunneling
regime.  These authors found by numerical calculation of the drift
velocity that for small electron-phonon coupling, there are
energies where the electron cannot propagate. In our approach,
this corresponds to the formation of the quasi-continuous band for
each of the WS ladders.  The electron cannot propagate when its
energy lies in the band gap.   However, with increasing coupling
constant, the band gap will disappear because of the overlap of
subsequent bands.  More quantitatively, let us consider the case
$t=\om=1$ (as the parameters in [\onlinecite{tru}]).  Since the
hopping constant is large, one needs to adopt the formalism in
this section. As seen in Fig.\ 13, the bandwidth increases with
increasing coupling constant, producing band overlap for large
enough $\gamma$.  This vanishing of the gaps as a function of
$\gamma$ would coincide with the resumption of particle drift in
the system. Just such behavior was found numerically in
[\onlinecite{tru}], and one can now explain it in terms of the
level structure of the system. It is also interesting to notice
that the overall envelope seen in $D_\nu$ is similar to that of
the drift velocity: small at the center and edge of each band, as
shown here in Fig.\ 7b, and qualitatively like the graphs in
figure 4 of reference [\onlinecite{tru}].

\begin{figure}
\includegraphics[width=3.7in]{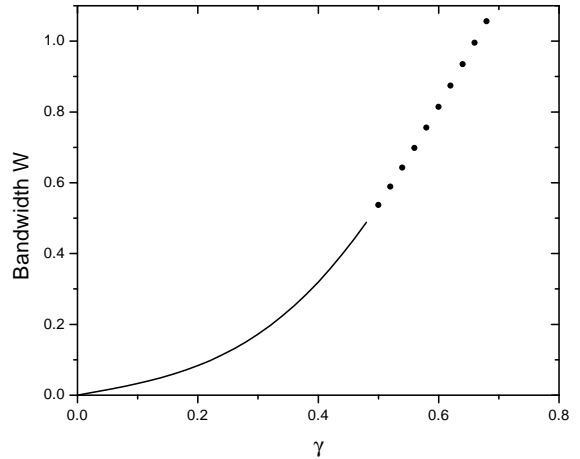}
\caption{ Bandwidth in the resonant and stronger coupling regime
as a function of coupling constant $\gamma$.  Here, $\De = \om = t
=1$. As $\gamma \agt 0.6$, neighboring bands will overlap,
indicated here with dotted line. }
 \label{fig13}
\end{figure}

\section{Conclusion} \label{discussion}

We have studied the coherent dynamics of Holstein polarons in a
strong electric field.  We have found that with the help of
phonons, a sort of quasi-continuous band will form under resonance
conditions, even for weak coupling constants. The band shows
approximate fractal self-similar structure, which is inherent in
the near self-similarity of the Hamiltonian. Although the phonons
can help the electron jump from one WS ladder state to another,
the phonons can also prevent the electron from propagating, if too
many phonons are involved.  These peculiar interaction differences
give rise to a variety of unusual states, including: (a)
delocalized polarons despite the strong electric field, with a
typical phonon cloud accompanying the electron; and (b) states
with high degeneracy at the band center, where the electron is
localized in a site of the lattice, and the phonon is located away
from the electron, in a stretched configuration. The band
structure is also manifested in transport properties of the
system, which we expect could be observed in tunneling
experiments. The level structure and extension of the different
states will appear as a modulation of the phonon replicas in the
tunneling experiment.  For weak coupling, this would only occur if
the system is in resonance, $\om = \De$, as away from that
condition the states are basically localized and would not
transport current (except for thermal effects). In a given
structure, the resonance condition can be reached by sweeping the
electric field, while monitoring the tunneling through the
structure.  As the spacing between the deformed rungs and the
localization length of the eigenfunction can be adjusted by
electric field, it would be quite interesting to see the
transition in experiments. With increasing coupling constant, the
`minibands' will overlap, and give rise to an overall merging of
the phonon replicas in tunneling, even when away from the
resonance regime.

 \begin{acknowledgments}
We thank A. Weichselbaum for help with graphics. This work was
supported in part by US DOE grant no.\ DE--FG02--91ER45334, and
the Condensed Matter and Surface Sciences Program at Ohio
University.  SEU appreciates the kind hospitality of K. Ensslin's
group during his stay at ETH.
 \end{acknowledgments}

\end{document}